\def\to{\hbox{$\,$--$\,$}}
\def\muspc{\hskip 0.15 em}

\def\mag{\hbox{$\;.\!\!\!^m$}}
\input psfig
\documentstyle{l-aac}
\begin{document}

\thesaurus{06(08.03.1, 08.16.4, 08.05.3, 08.08.1, 11.09.1
Sagittarius dwarf; 11.12.1)}

\title{Are the Bulge C{\to}stars in the 
Sagittarius dwarf galaxy?}

\author{Yuen K. Ng}

\institute{Padova Astronomical Observatory,
           Vicolo dell'Osservatorio 5, I-35122 Padua, Italy 
           ({\tt Yuen\char64astrpd.pd.astro.it})}
\date{Received 21 March 1997 / Accepted 10 July 1997}

\maketitle

\markboth{Yuen K. Ng:\ Are the Bulge carbon stars in the
the Sagittarius dwarf galaxy?}{}

\begin{abstract}
Part of the mystery around the Bulge carbon
stars from Azzopardi et al. (1991) is solved, if
they are related to the Sagittarius dwarf galaxy. The carbon stars are 
in that case
not metal-rich as previously thought, but they have a metallicity
comparable to the LMC, with an age between 0.1\to1~Gyr.
A significant fraction of the carbon stars still
have luminosities fainter than the lower LMC limit
of \mbox{M$_{\rm bol}\!\simeq$\muspc\to3\mag5}.
A similar trend is present
among some of the carbon stars found in other dwarf spheroidals,
but they do not reach a limit as faint as 
\mbox{M$_{\rm bol}\!\simeq$\muspc\to1\mag4} found for the SMC.
At present, the TP-AGB models cannot explain the origin 
of carbon stars with \mbox{M$_{\rm bol}\!>$\muspc\to3\mag5}
through a single-star evolution scenario,
even if they form immediately after entering the
TP-AGB phase.
Mass transfer through binary evolution is suggested 
as a possible scenario to explain the origin of these low
luminosity carbon stars.
\end{abstract}

\section{Introduction}
\subsection{Bulge carbon stars}
For a long time carbon stars are searched
in the direction of the Galactic Centre.
From a low dispersion, near infrared
grism survey 
a total of five carbon stars were found amid
2187\muspc M-giants (Blanco et al. 1978, McCarthy et al. 1983 and
Blanco \& Terndrup 1989).
The stars are mainly identified by the
CN bands at 7945, 8125, and 8320 \AA.
Azzopardi et al. (1985ab; 1986) demonstrated that additional, especially
blue carbon stars can be found 
with the strong Swan C$_2$ bands
(4737 and 5165 \AA) in the spectral range 4350\to5300 \AA\,.
Using this technique,
Azzopardi et al. (1985b, 1988; the latter
is hereafter referred to as ALR88) found 33 carbon stars
in 8 different fields of the Galactic Bulge.
Near-IR photometry and medium-low resolution spectra have been
obtained for these carbon stars
(Azzopardi et al. 1991 -- hereafter referred to as ALRW91, 
Tyson \& Rich 1991, Westerlund et al. 1991).
These stars show similarities with the low- to medium bolometric
luminosity SMC carbon stars, but the
galactic carbon stars have stronger NaD doublets.
Various studies suggest that a wide metallicity range
is present in the bulge
(Whitford \& Rich 1983,
Rich 1988 \& 1990, Geisler et al. 1992, McWilliam \& Rich 1994,
Ng 1994, Bertelli et al. 1995, Ng et al. 1995 \& 1996, Sadler et al.
1996). 
According to ALR88 the carbon stars 
are expected to be metal-rich 
if they belong to the Bulge.

\subsection{Sagittarius dwarf galaxy}
The serendipitous identification of the 
Sagittarius dwarf galaxy (SDG) was made by Ibata et al. (1994, 1995;
hereafter respectively referred to as IGI94 and IGI95).
It is the closest dwarf spheroidal and moves away from us 
at about 160~km/s. 
Accurate distance determinations from RR Lyrae stars belonging to this galaxy
range from 22.0\to27.3~kpc (Alcock et al. 1997; Alard 1996; 
Mateo et al. 1995ab, 1996; Ng \& Schultheis 1997 -- hereafter referred
to as NS97).
The photometric metallicity estimates made thus far depend heavily
on the assumed age and the values for [Fe/H] range from
--0.5 to --1.8. The mean age (10\to12~Gyr) and metallicity 
([Fe/H]\muspc=\muspc\to1.5) adopted is a trade-off, such that
the age conveniently allows for the presence RR Lyrae and carbon stars
(Ibata et al. 1997).
IGI95 identified four carbon stars belonging 
to this galaxy, Whitelock et al. (1996, hereafter referred to as WIC96) 
performed a near-IR study of 26 candidates, and NS97
identified one more carbon star in the outer edge of the dwarf 
galaxy. The next section deals with how the Bulge carbon stars 
mentioned above are related to the 
(candidate) carbon stars from the dwarf galaxy.

\subsection{A clue?}
A supposedly high metallicity lead 
Tyson{\muspc\&\muspc}Rich (1991) and Westerlund et al. (1991) to suggest
that the Bulge carbon stars should be old
and posses a mass of about 0.8~M$_\odot$,
while evolutionary calculations (Boothroyd et al. 1993, 
Groenewegen \& de Jong 1993, Groenewegen et al. 1995, 
Marigo et al. 1996ab) demonstrate that the initial mass 
of carbon stars is at least 1.2 M$_\odot$ ($t\!\la\!4$~Gyr)
for \mbox{Z\muspc=\muspc0.008}. Furthermore, the initial mass increases 
towards higher metallicity and decreases towards lower metallicity
(Lattanzio 1989).
The Bulge carbon stars are a mystery 
(Lequeux 1990, Tyson{\muspc\&\muspc}Rich 1991, 
Westerlund et al. 1991, Chiosi et al. 1992,
Azzopardi 1994), because they 
are in bolometric luminosity about 2\mag5 too faint
to be regarded as genuine AGB (Asymptotic Giant Branch) stars, 
if located in the metal-rich Bulge.
\par
NS97 compiled a catalogue of candidate RR lyrae and 
long period variables (LPVs) in the outer edge of the SDG.
After de-reddening of the K-band magnitudes for 
the candidate LPVs, the  
period-K$_0$ relation for the Mira variables 
from the Sgr~I field in the Bulge from Glass et al. (1995)
put these stars at 25.7~kpc. This distance is in 
good agreement with the distance obtained from the RR~Lyrae 
stars found in the same field. 
One of the LPVs is a carbon star.
The difference between the distance modulus of the dwarf 
galaxy and the Galactic Centre at 8~kpc is $\sim$2\mag5.
This lead to the suggestion that the Bulge carbon stars
could actually be located in the dwarf galaxy, whose presence
was unknown at the time the carbon stars were identified.
This would solve the standing question about the origin 
of the `bulge' carbon stars.
\par
The possibility that the `bulge' carbon stars could be member
of the SDG is analyzed.
The organization of the paper is as follows. In Sect.~2
the formerly Bulge carbon from ALRW91 are placed 
in a CMD (colour-magnitude diagram) at the 
distance of the SDG. A comparison is made with the
near-IR magnitudes and colours from actual and candidate  
carbon star members from the dwarf galaxy (WIC96, NS97).
In Sect.~3 isochrones are placed
in the CMD and it is demonstrated that the majority of the ALRW91 carbon
stars are still fainter than the tip of the red giant branch.
A discussion of the results is given in
Sect.~4. Some stars are too faint, even if carbon stars are formed
immediately after they enter the TP-AGB phase,
and a possible binary evolution origin 
is suggested for some of the carbon stars 
fainter than the red giant branch tip.
Arguments are given that the expected number of carbon stars 
related to the SDG is at least two times larger.
The results are summarized in Sect.~5.

\section{Data and method}

\subsection{Near-infrared photometry}
The near-IR photometry of the Bulge carbon stars
presented by ALRW91 was reduced to
the standards from the homogenized ESO photometric system 
proposed by Koornneef et al (1983ab). 
Bouchet et al. (1991) found in this system no systematic differences
in the magnitudes from the standards observed at ESO 
in the period 1983\to1989.
\par
\begin{figure}
\centerline{\vbox{\psfig{file=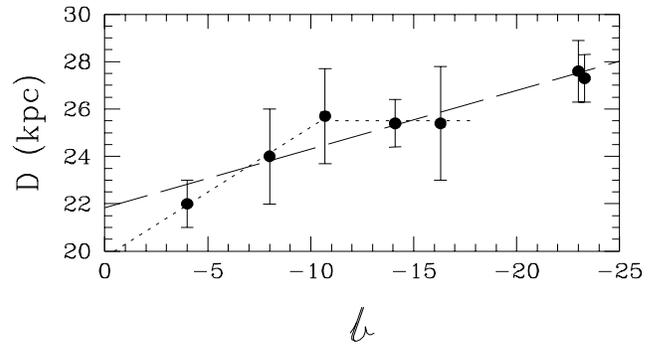,height=4.21cm,width=8.5cm}}}
\caption{The distance of the Sagittarius dwarf galaxy determined for various
galactic latitudes. The dashed line refers to an unweighted linear least
squares fit for the points, while the long-dashed line shows
a two section fit drawn through the points}
\end{figure}
The near-IR photometry from WIC96
with the SAAO 1.9m telescope is appropriately 
transformed to the ESO system (Hron et al. 1997).
Note, that in general many transformations from SAAO to the ESO near-IR 
system refer to the transformation from the system
defined by Glass (1974) or Carter (1990) to an older ESO 
system defined by Engels et al. (1981) and Wamsteker (1981).
\par

\subsection{Extinction and distance}
The extinction for the Bulge carbon stars is not
homogeneous. For each field defined by ALRW91
a general correction is applied to all stars.
Instead of \mbox{A$_V$\muspc=\muspc1\mag87} (Glass et al. 1995)
an extinction of \mbox{A$_V$\muspc=\muspc1\mag71} is adopted 
for the Sgr~I field. It is the average 
value from Walker\mbox{\muspc\&\muspc}Mack (1986) 
and Terndrup et al. (1990).
For Baade's Window field around NGC~6522  
\mbox{A$_V$\muspc=\muspc1\mag50}. This value 
was obtained by Ng et al. (1996) from the
(V,V--I) CMD obtained by the OGLE (Optical Gravitational Lensing Experiment,
see Paczy\'nski et al. 1994 for details)
and is in good agreement with average value
\mbox{E(B--V)\muspc=\muspc0\mag49} determined for this field.
For the extended clear region around and near NGC~6558 the same value
as used by ALR88, 
\mbox{E(B--V)\muspc=\muspc0\mag41} or 
\mbox{A$_V$\muspc=\muspc1\mag27} from Zinn (1980), is adopted.
No actual determinations of the extinction are found 
for the field intermediate to NGC~6522 and NGC~6558
and \mbox{A$_V$\muspc=\muspc1\mag38} is adopted.
For the remaining Bulge fields ($b\!<\!-8\fdg0$) 
a relation (Blommaert 1992, Schultheis et al. 1997)
based on the reddening map constructed by
Wesselink (1987) from the colour excess of the RR~Lyrae stars
at minimum light, is used: 
${\rm A}_{V}\!=\!0.104\;b+1.58$, where $b$ is the
galactic latitude.
\hfill\break
The value \mbox{A$_V$\muspc=\muspc0\mag48} 
obtained by Mateo et al. (1995) is adopted
for the stars in the SDG observed by WIC96.
The extinction in the near-IR passbands is determined 
under the assumption that 
\mbox{A$_J$/A$_V$\muspc=\muspc0.282},
\mbox{A$_H$/A$_V$\muspc=\muspc0.175}, and
\mbox{A$_K$/A$_V$\muspc=\muspc0.112} (Rieke \& Lebofsky 1985).
\par
The distance determined with RR Lyrae stars 
for the SDG ranges from \mbox{22.0\to27.3~kpc}.
Figure~1 shows that these distances are actually correlated with 
the galactic latitude at which they were determined. Additional distance
determinations (Sarajedini\&Layden 1995 -- hereafter referred to as SL95, 
Fahlman et al. 1996)
obtained with other methods were added to this figure.
An unweighted linear least squares fit through those distances 
gives $D{\rm(kpc)}\!=\!21.83-0.25\;b $. Figure~1 also shows a two 
section line drawn through these points. 
\begin{figure*}
\centerline{\vbox{\psfig{file=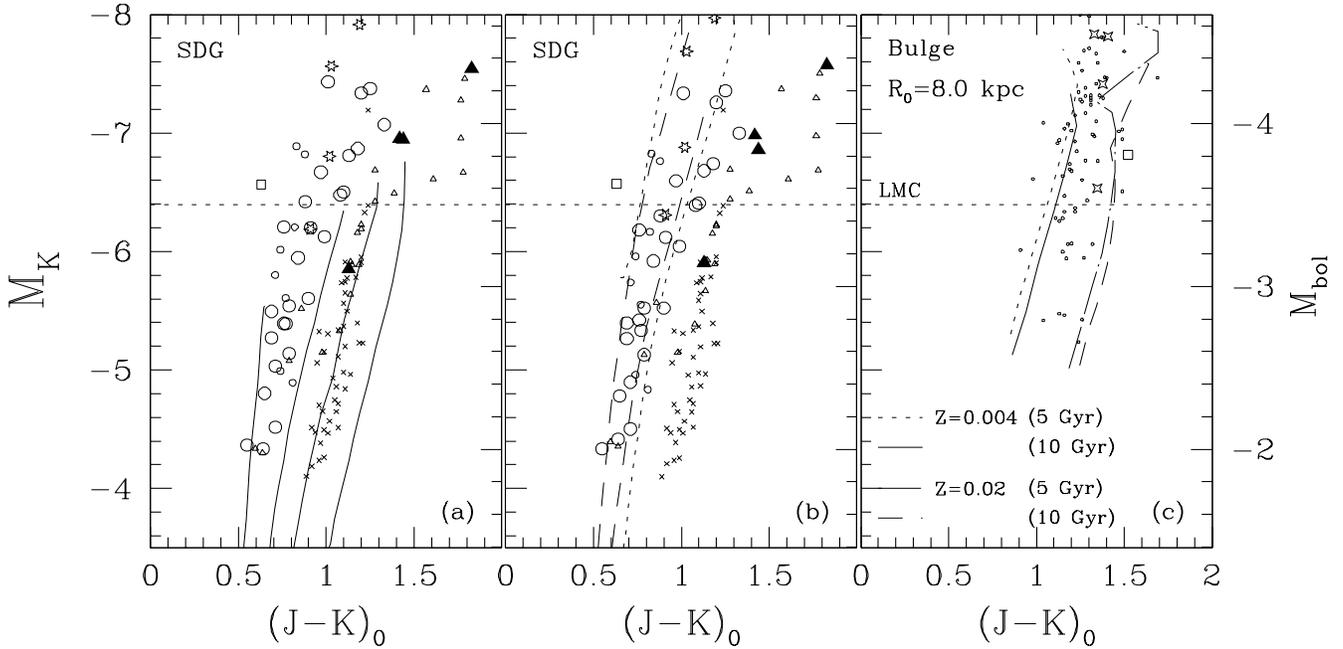,height=8.5cm,width=17.5cm}}}
\caption{Panel ({\bf a\muspc\&\muspc{b}})
The ALRW91 carbon stars placed at the distance of the Sagittarius 
dwarf galaxy (open circles; small open circles are used to indicate
a larger uncertainty in the extinction adopted), together with
the carbon stars from IGI95 (filled triangles).
Open triangles and crosses are used for respectively 
the candidate carbon stars and giant branch stars of the dwarf galaxy observed 
by WIC96. The open square indicates the carbon stars S283 found by NS97
among the 
variables studied by Plaut (1971) and Wesselink (1987).
The dotted horizontal line indicates the 
observational lower LMC limit at which carbon stars are found,
while the lower magnitude limit corresponds with the SMC limit
(Azzopardi 1994).
The difference between panel ({\bf a}) and ({\bf b}) is the method
used to assign a distance to each star, see Sect.~2.3.
Panel ({\bf c}) shows the stars which are possibly located in the
Bulge. The open crosses indicate the giant branch stars observed by 
WIC96, the open square is a carbon star (L199, unpublished) found 
among the variables studied by Plaut (1971) and Wesselink (1987),
and the dots show for comparison the location of 
the semiregular and Mira variables selected from
respectively Schultheis et al. (1997) and Blommaert (1992).
\newline
The isochrones (Bertelli et al. 1994; only for log~$T_{\rm eff}\!<\!3.70$) 
displayed in the three 
panels are as follows:
{\bf a}) 10 Gyr RGBs for
Z\muspc=\muspc0.0004, 0.004,0.008, and 0.020;
{\bf b}) 1 and 10 Gyr AGBs for \mbox{Z\muspc=\muspc0.001} (long-dashed line)
and 0.1 and 1 Gyr AGBs for \mbox{Z\muspc=\muspc0.008}
(dotted line); and
{\bf c}) 5 and 10 Gyr AGBs for \mbox{Z\muspc=\muspc0.004}
(dotted and solid line) and \mbox{Z\muspc=\muspc0.02}
(dot-dashed and dashed line).
See Sect.~3.1 for additional details about the 
magnitude and the colour transformations.
\newline
Note that the left vertical axes in each panel 
give the K-magnitude scale, while the right vertical axes 
indicate the M$_{bol}$-scale}
\end{figure*}

\subsection{Colour-magnitude diagram}
Figure 2a shows the de-reddened CMD. A distance correction with the
linear least squares line was applied to all the stars to reduce 
the scatter due to differential distance effects. Figure~2b
shows the diagram with the distances corrected with the two section line.
There is no significant difference between the Figs.~2a\muspc\&\muspc2b.
Some stars are very bright. In SDG their M$_{bol}$ would range from 
\mbox{\to6\mag0} to \mbox{\to7\mag0}.
They could be bright members of SDG or they could be located in the
Bulge and have a lower luminosity. For a graphical purpose  
those stars are placed in Fig.~2c at a distance of 8.0~kpc. 
A sample of LPVs from Schultheis et al. (1997) are included
in panel~2c.
The dotted line in Figs.~2a\to{c} is the lower LMC limit at which
carbon stars are found,
while the lower magnitude boundary corresponds with the SMC limit. 
\par
Figures~2a\muspc\&{b} show that three of the four IGI95 carbon stars
observed by WIC96 form an extension to the ALRW91 sequence 
of carbon stars. This clearly is a strong indication that the ALRW91 
carbon stars belong to the SDG. 
The fourth carbon star (star C1 from WIC96) appears to be redder and located
in the parallel sequence formed by the giant branch stars. 
It is not clear if this is due to a mis-identification in the photometry 
with another redder, nearby star. The finding chart provided from WIC96  
does not rule out this possibility.
The figure further shows that
eight (\#s 1\to3,\muspc5,\muspc6,\muspc12,\muspc15,\muspc21) 
out of the 26 stars from WIC96 are highly eligible carbon star candidates,
because they are located on the combined ALRW91 \& WIC96 carbon star sequence. 
\par
A significant number of the ALRW91 carbon stars are located below
the LMC limit at which carbon stars are found.
This limit might be related to a low metallicity of the stars,
but in Sect.~4.1 it is argued that this is not the case.
A similar trend is present among some of the carbon stars found in other 
dwarf spheroidals (see Fig.~2b from WIC96).
However, Azzopardi et al. (1997) found carbon stars with 
\mbox{M$_{bol}\!\simeq$\muspc\to1\mag2}
in the Fornax dwarf galaxy, assuming 
\mbox{(m--M)$_0$\muspc=\muspc21\mag0}.
This is even fainter than the present limit 
\mbox{M$_{bol}\!\simeq$\muspc\to2\mag0} for the `bulge' carbon stars
if located in the SDG and 
\mbox{M$_{bol}\!\simeq$\muspc\to1\mag4} for the SMC 
carbon stars (Azzopardi 1994; Westerlund et al. 1993, 1995).
\hfill\break
Four LPVs, most likely belonging to 
the dwarf galaxy (NS97), are included in this figure. 
Note that they are the first Mira and the semiregular variables found  
belonging to the SDG. In fact, they 
are the first LPVs discovered in a dwarf spheroidal galaxy.
Two LPVs are located inside the carbon star sequence, while the
other two form at \mbox{M$_K\!<$\muspc\to7\mag5}
a blue extension to the carbon sequence. 

\section{Models}
\subsection{Conversion: theoretical $\longrightarrow$ observational plane}
%\subsection{M$_{bol} \longrightarrow$\muspc{M}$_K$
%and $\log T_{\rm eff} \longrightarrow$\ (J--K)$_0$}
The transformation of the bolometric magnitude to the K-magnitude
scale in Fig.~2 is obtained from the period-luminosity (PL- and PK-) relation
of carbon Miras (Groenewegen{\muspc\&\muspc}Whitelock 1996,
hereafter referred to as GW96).
Transformed to the ESO photometric system this gives:
\mbox{M$_K$\muspc=\muspc1.37M$_{bol}$\to1.58}.
The actual transformation from the bolometric magnitude 
to a K-magnitude in the models shown in Fig.~2 
is achieved with the relation provided by 
Suntzeff et al. (1993): 
\begin{eqnarray*}
\quad &{\rm M}_K&={\rm M}_{bol}-BC_K \;, \\
\quad &BC_K&=1.00+2.076({\rm J}-{\rm K})_0-0.463({\rm J}-{\rm K})_0^2 \;. 
\end{eqnarray*}
The relation is valid for 
\muspc\mbox{0\mag3\muspc$<\!({\rm J}-{\rm K})\!<$\muspc2\mag3}\muspc
and was obtained from a fit to the (J--K,$BC_K$) relation
given by Frogel et al. (1980). 
%$BC_K$ is the difference between the K magnitude and 
The bolometric magnitude was obtained from the numerical integration of the 
broadband flux distributions of the galactic carbon stars
observed by Mendoza \& Johnson (1965). 
The same bolometric corrections are applied to non-carbon stars. Figure~2 
from Frogel et al. (1980) shows that the differences are at most 
$\sim$\muspc0\mag2. 
\par
The conversion from $\log T_{\rm eff}$ to (J--K)$_0$ is established 
from an empirical relation derived by Ng et al. (1997), based on 
giant stars with a spectral type ranging from late G to late M.
The effective temperatures for these stars 
were obtained from angular diameter measurements.
The empirical $\log T_{\rm eff}$/(J--K)$_0$
relation takes into account a small shift in colour as a function 
of metallicity. 
The relations, provided for the metallicity \mbox{Z\muspc=\muspc0.004} 
and \mbox{Z\muspc=\muspc0.024}, are logarithmically interpolated
for \mbox{Z\muspc$\ge$\muspc0.004} and the \mbox{Z\muspc=\muspc0.004}
relation is applied for lower metallicities.
Although some uncertainty is present due to the fact that this relation is not 
based on carbon stars, the differences will not
be too large as long as the colours are not too red.
The empirical relation covers conveniently the 
observed colour range of the carbon stars. 

\subsection{Isochrones}
In Figs.~2a\to{c} some isochrones from Bertelli et al. (1994)
are displayed. Fig.~2a shows the RGB for an age of 10 Gyr
for various metallicities and Fig.~2b\muspc\&\muspc2c
show the AGB for different age and metallicities.
Figure~2a demonstrates that if one assumes a fixed age then
a large metallicity spread could be present among the SDG 
stars. Given the uncertainties, 
the GB stars likely have a metallicity 
of \mbox{Z\muspc=\muspc0.008}, comparable to the LMC.
%The GB stars with M$_K$ in the range between \mbox{\to4\mag0}
%and \mbox{\to5\mag0} are likely even older GB stars.
Figure~2b demonstrates that difficulties are present to distinguish
1 \& 10~Gyr isochrones for the AGBs with
\mbox{Z\muspc=\muspc0.001} from those 
with a 0.1 \& 1~gyr age for \mbox{Z\muspc=\muspc0.008}.
The isochrones in Fig~2c show that the colour difference
between 5 and 10~Gyr populations with the same metallicity is quite small
compared to the colour difference due to a large metallicity range.

\section{Discussion}

\subsection{Metallicity and age}
The isochrones in Fig.~2a indicate that the metallicity of the old
population in the SDG is 
\mbox{Z\muspc$\simeq$\muspc0.008}. This is quite high, but 
within the uncertainties comparable with 
\mbox{[Fe/H]\muspc=\muspc\to0.52$\pm$0.09} obtained by
SL95 from a photometric study of a (V,V--I) CMD.
This is in contrast with values \mbox{[Fe/H]\muspc$\la$\muspc\to0.7}
obtained with different methods (IGI94, IGI95, Ibata et al. 1997,
Fahlman et al. 1996, Mateo et al. 1995 \& 1996, WIC96, Marconi et al. 1997).
\hfill\break
SL95 found an indication of the possible existence of a 
\mbox{[Fe/H]\muspc=\muspc\to1.3} (i.e. \mbox{Z\muspc=\muspc0.001})
component in the dwarf galaxy.
With such a metallicity the age of the carbon stars
is somewhere between 1 and 10~Gyr, see Fig.~2b.
Figure~2a already shows that there is an old component 
with \mbox{Z\muspc=\muspc0.008} and an age around 10~Gyr.
If there is a rather smooth age range present for 
\mbox{Z\muspc=\muspc0.001}, then why are the younger
stars absent among the RGB stars with \mbox{Z\muspc=\muspc0.008}\muspc?
Furthermore, why should the metallicity decreases towards 
younger age\muspc? From a close box model one expects
an increasing metallicity towards younger age.
All together this does not favour a metallicity of 
\mbox{Z\muspc=\muspc0.001}.
\par
\begin{figure}
\vbox{\null
\centerline{\psfig{file=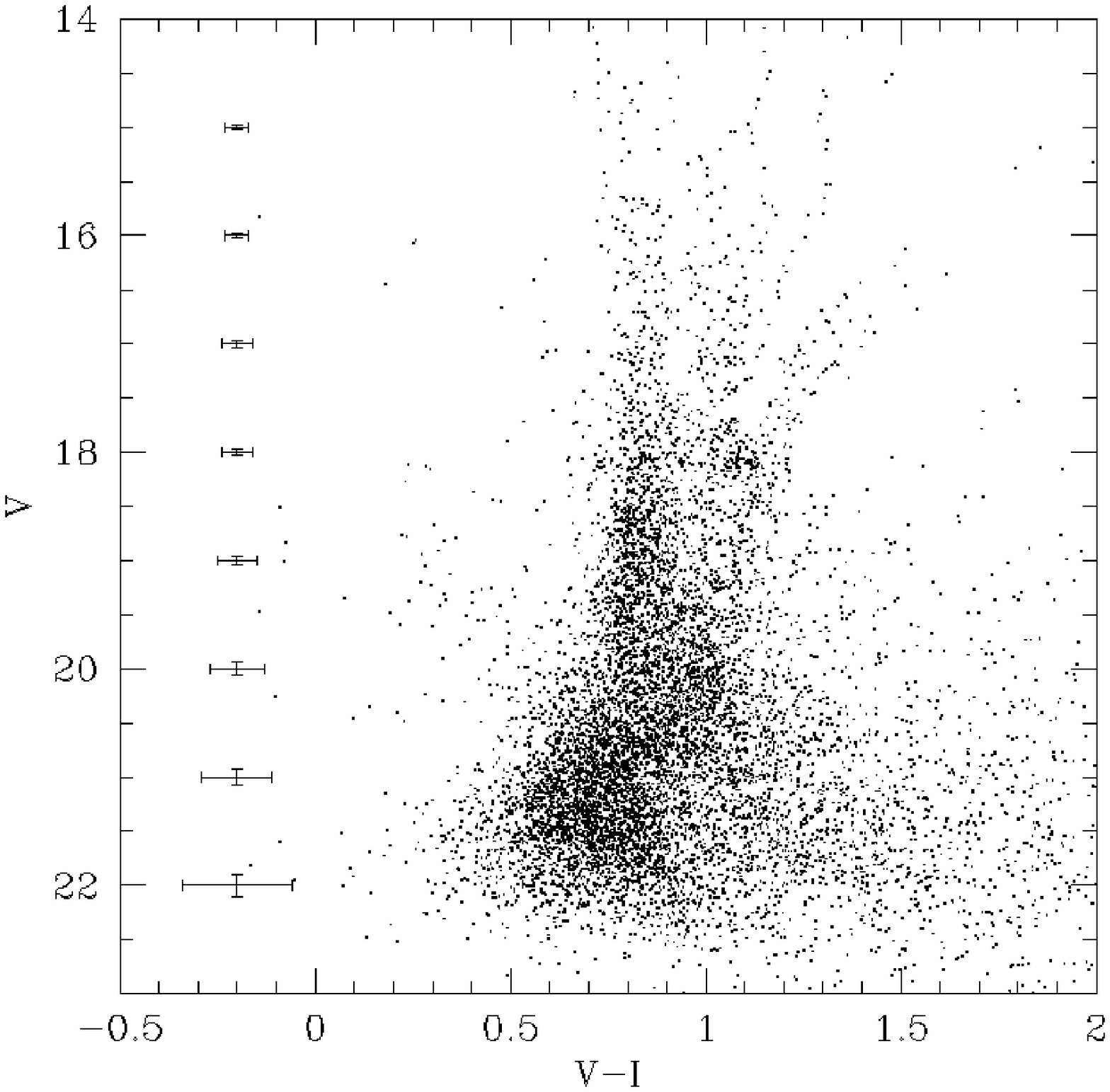,height=10.3cm,width=8.5cm}}
\vskip-10.37cm
\centerline{\psfig{file=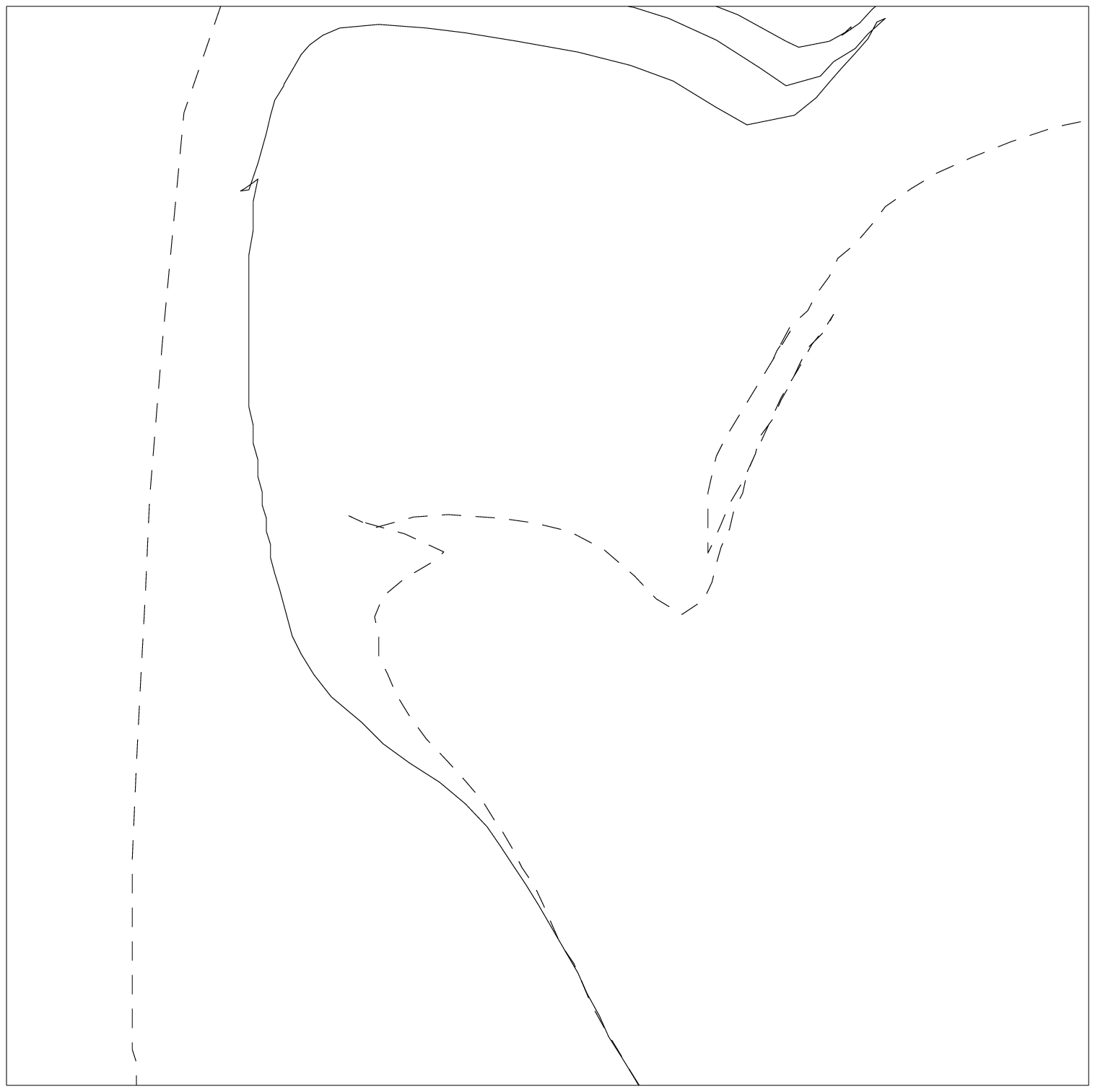,height=9.14cm,width=9.675cm}\hskip0.05cm\null}
\vskip -1.0cm
}
\caption{CMD of SDG stars in field~1 from Marconi et al. (1997)
with an overlay of \mbox{Z\muspc=\muspc0.008} isochrones of 
0.1~Gyr (solid line) and 1~Gyr (dashed line) from Bertelli et al. (1994). 
The dashed line between \mbox{(V,V--I)\muspc=\muspc(14\mag0,0\mag0)}
and \mbox{(V,V--I)\muspc=\muspc(23\mag0,\to0\mag2)}
is the white dwarf cooling sequence from the 1~Gyr isochrone.
An extinction of \mbox{A$_V$\muspc=\muspc0\mag48}
and a distance modulus of 17\mag02 (Mateo et al. 1995)
have been adopted for the isochrones}
\end{figure}
\noindent
If on the other hand, the metallicity for the carbon stars 
is \mbox{Z\muspc=\muspc0.008} then we are apparently 
dealing with two different age populations.
The carbon stars would in this case reflect a very 
recent star formation burst in this galaxy,
while the major stellar population has an age \mbox{$\sim$\muspc10~Gyr}.
An almost zero metallicity enrichment is 
in agreement with the multiple starburst Carina dwarf 
spheroidal studied by Schmecker-Hane et al. (1996).
\hfill\break
A young age could be independently confirmed through the 
detection of Cepheids in the SDG. Mateo et al. (1995) report
the possible detection of an anomalous Cepheid and suggest that
the SDG might contain a considerable number of these stars.
\par
\begin{figure}
\centerline{\vbox{\psfig{file=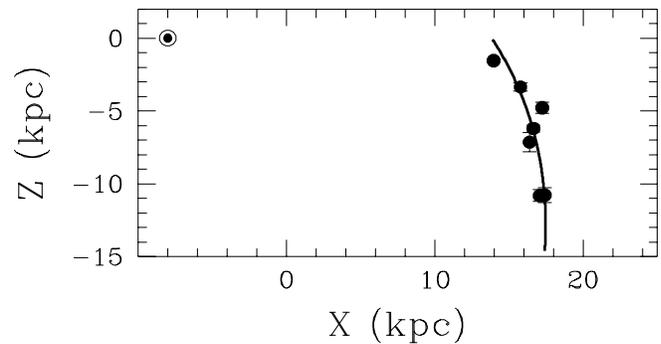,height=4.5cm,width=8.5cm}}}
\caption{The course of the Sagittarius dwarf galaxy projected on 
the $x-z$ plane. The radial velocity of the galaxy indicates 
that it is moving away from us. Together with the indicated course
this implies that the galaxy crossed quite recently the galactic disc.
The position of the Sun is indicated by $\odot$}
\end{figure}
Just as in this study, SL95 would not be able to distinguish
from their CMD the difference between a very young 
\mbox{Z\muspc=\muspc0.008} population from an older 
\mbox{Z\muspc=\muspc0.001} population. So the present interpretation
is not in contradiction with their results. 
An indication that we are indeed dealing with young stellar populations
is present in the CMDs from
Marconi et al. (1997). In Fig.~3 (i.e. Marconi's et al. Fig.~2)
the 0.1~Gyr and 1.0~Gyr isochrones demonstrate clearly the presence of
a young population above the main sequence turn-off of 
the old stellar population from the SDG.
It is not clear though, why the two 
distinct age populations advocated above are not present 
in the CMDs analyzed by Fahlman et al. (1996).
Possibly the number of stars involved are
too small to be conclusive and the analysis has to be
repeated with a larger number of stars in the background 
of the globular clusters.
\par
From the considerations outlined above it is concluded 
that the metallicity of the SDG stars is 
\mbox{Z\muspc$\simeq$\muspc0.008}. The age of the 
ALRW91 carbon stars is with this metallicity younger than 
1~Gyr. A second distinct population is present 
with an age around 10~Gyr.
\hfill\break
If a star formation burst is related with 
a passage through the galactic disc then 
an age as young as 0.1~Gyr would confirm the assertion
that the SDG already passed through the disc 
and is currently moving away from our Galaxy
(Alcock et al. 1997, NS97).
Figure~4 indicates that this should indeed be the case.
The absence of populations 
with intermediate ages separated by approximately 1~Gyr
appears to rule out the orbital period suggested
by Johnston et al. (1995), Vel\'azquez{\muspc\&\muspc}White (1995) and 
Ibata et al. (1997).
Johnston et al. however pointed out that the existing observations 
were not sufficient to put limits on the orbit or initial state
of the SDG. 

\subsection{Are the `bulge' carbon stars on the TP-AGB ?}
In the current understanding of stellar evolution theory carbon stars
are formed at the third dredge-up at the TP-AGB 
(thermally pulsing asymptotic giant branch)
phase, in which the stellar surface is enriched with $^{\rm 12}$C.  
Carbon stars cannot be formed before the TP-AGB phase.
In the LMC the majority of the stars are
located above the tip of the RGB (red giant branch),
but they can be located \mbox{$\sim$\muspc1\mag0} below RGB tip
for 20\%\to30\% of its interpulse period (see Marigo et al. 1996ab 
and references cited therein).
\par
\begin{figure}
\centerline{\vbox{\psfig{file=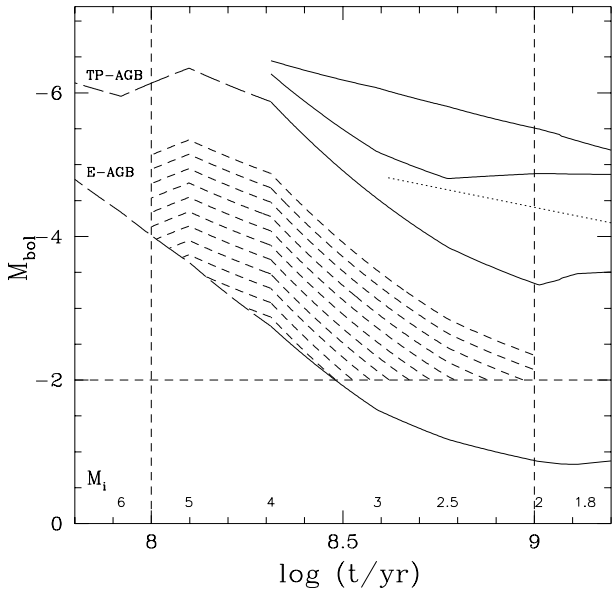,height=8.2cm,width=8.5cm}}}
\caption{Four solid lines are displayed for the LMC, 
indicating the start of the E-AGB, TP-AGB,
the transition from M- to C-type stars and 
the end of the AGB evolution (see Marigo et al. 1996b for details). 
Above the dotted line carbon stars with masses in the range
1.2\to3.0~M$_\odot$ have a 90\% detection probability.
M$_{\rm i}$ is the initial mass of the carbon star.
The dashed, horizontal line is the limiting magnitude for
the SDG carbon stars and the dashed, vertical lines indicate
the age range for the SDG stars (see Sect.~4.1) if 
\mbox{Z\muspc$\simeq$\muspc0.008}. 
The shaded area indicate the region in which the presence 
of carbon stars cannot be explained with a TP-AGB star,
see Sect.~4.2
}
\end{figure}
At the distance of the SDG 
a considerable fraction of the ALRW91 carbon stars 
are located well below the tip of the RGB. 
Figure~5 demonstrates that
even if carbon stars can form directly after they enter 
the TP-AGB phase, the presence of a fraction
of those carbon stars cannot be explained.
These stars are likely formed through another mechanism 
not accounted for in the `standard theory' for the formation
of carbon stars in the TP-AGB phase.
The sequence of carbon stars parallel to the giant branch
suggest that they are either RGB or early AGB stars. It
apparently contradicts our theoretical understanding about the
formation process of carbon stars.
\par
ALR88 suggested two possible scenarios. Westerlund et al. (1991) 
favoured the scenario where a very effective mixing occurs 
at an early phase on the ascend to the TP-AGB. 
Since this cannot be the case for even the faintest
carbon stars related to the SDG, the scenario of
mass transfer through binary evolution is favoured
for the formation of the low luminosity carbon stars.
The same scenario explains the observations of
dwarf carbon stars (see Green 1997 and references cited therein).
\par
In case of mass transfer through binary evolution, 
the primary star evolved through
the TP-AGB phase and transferred a significant part of its
$^{\rm 12}$C enriched envelope to the secondary. The 
primary evolved away from the AGB and is a white dwarf now and
the secondary is the presently observed carbon star. 
The fully convective envelope of stars at the RGB or AGB would dilute
$^{\rm 12}$C/$^{\rm 13}$C ratio.
The dilution is proportional to the time elapsed since the primary 
deposited part of its envelope on the secondary star.
The dilution should increase towards larger distances from 
the galactic disc (see Figs.~\mbox{1\muspc\&\muspc4}).
The NaD equivalent width and the CN line strength obtained from 
spectra for the carbon stars by Tyson\mbox{\muspc\&\muspc}Rich
(1991, see their Fig.~3) would support this assertion. 
For comparable envelope masses the dilution will be at least a factor 2.
If the metallicity is about {Z\muspc=\muspc0.008} then the mass transfer
should have occurred recently, indicating that the difference
of the initial mass between the primary and secondary is very small.
A metallicity of \mbox{Z\muspc=\muspc0.008} is
in addition high enough to explain the strong NaD features
found present among the carbon stars.
\par
McClure (1984) showed that CH stars are binaries,
some of them are long-period systems.
Suntzeff et al. (1993) argue that the CH stars in the LMC,
with M$_{bol}$ ranging from \mbox{\to5\mag3} to \mbox{\to6\mag5},  
have an age near 0.1~Gyr. Some of the CH stars could well 
be the predecessors of the SDG carbon stars. 
If L199 is a CH star and the four giant stars observed by WIC96 are 
related to the SDG they have 
\mbox{M$_{bol}$\muspc$\simeq$\muspc6\mag5}.
A binary nature of those stars
would strongly support the evolution scenario outlined above.
\par
In Fig.~6 the near-infrared photometry of the 
`bulge' carbon stars are compared with the photometry for the 
obtained for a sample of SMC carbon stars (Westerlund et al. 1995).
For the SMC carbon stars we adopted 
\mbox{(m--M)$_0$\muspc=\muspc18\mag9} and 
\mbox{E(B--V)\muspc=\muspc0\mag09}
(Westerlund 1997).
Westerlund et al. (1991, 1993, 1995) noted that the most luminous 
`bulge' and SMC carbon stars have comparable C$_2$ and CN values,
whereas the fainter `bulge' stars are more similar to the main bulk of
faint SMC carbon stars. This result is not surprising anymore
if, as argued in this paper, the `bulge' carbon stars are indeed related  
to the SDG. Figure~6 indicates that the former `bulge' carbon stars have
luminosities almost comparable to the SMC carbon stars.
The ALRW91 carbon stars form a parallel sequence with the SMC stars.
As argued in Sect.~4.1 the ALRW91 carbon stars likely have a
metallicity \mbox{Z\muspc$\simeq$\muspc0.008}, while the SMC carbon 
stars are expected to be metal poorer. The SMC sequence ought to be bluer
than the SDG sequence, if the carbon stars have a similar age. 
This however is not the case and the carbon stars therefore do not
originate from a population with a similar age. The colours of both sequences 
can be explained when the SMC carbon stars evolved from an older stellar 
population. 
\par
\begin{figure}
\centerline{\vbox{\psfig{file=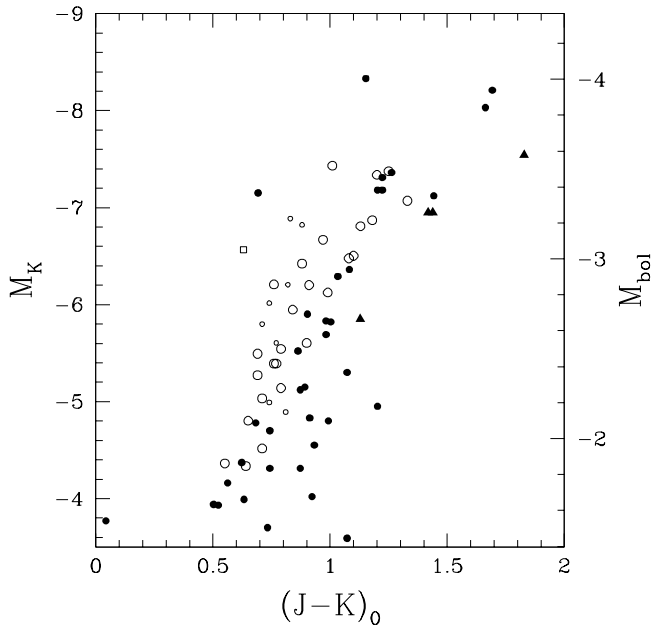,height=8.2cm,width=8.5cm}}}
\caption{A comparison between the `bulge' carbon stars (symbols
as in Fig.~2) if located in the SDG and the SMC carbon stars
(Westerlund et al. 1995; filled dots)}
\end{figure}

\subsection{Are there more SDG carbon stars ?}
It is an important issue to check if more carbon stars from the dwarf
galaxy can be found in other fields. From NS97 a crude estimate
of about 1/2 is found for the ratio of LPVs 
versus the RR Lyrae stars of Bailey type {\it ab}. 
About one out of six LPVs 
could be a carbon star. This is however a lower limit, because the sample
from NS97 is biased towards the brighter stars. 
The observations were mainly limited due to the instrumental setup
and the telescope size and no reliable data was obtained for
some of the fainter, suspected carbon stars in the sample.
\par
About 29 more carbon stars with \mbox{M$_K\!\la$\muspc\to6\mag5}
might be found among the long period variables in the 
DUO (Disc Unseen Objects; see Alard 1996 and references cited therein),
MACHO (Massive Astrophysical Compact Halo Objects; see references
in Alcock 1997), and even the OGLE (see references 
in Paczy\'nski et al. 1994) databases.
However, considerably more candidate carbon might be present among the 
stars with \mbox{\to6\mag5\muspc$<\!{\rm M}_K\!<$\muspc\to4\mag0}.
Extrapolating the carbon star PK-relation from GW96 indicates
that if some of these stars are variables, 
they might be found among the variables with periods
in the range from 40\to130 days.
An additional result would be a significant increase of the number
of LPVs related to the SDG.
\par
Another way to estimate the expected number of carbon stars for the
SDG is through the fuel consumption theorem
(see Renzini\mbox{\muspc\&\muspc}Buzzoni 1983, 1986 for details).
For a population of young stars we have 
the relation: $N_*\!=\!5\times10^{-12}\times\Delta{t}\times{L_{bol}}$,
where $L_{bol}\!\sim\!4\!\times\!10^{7} \ L_\odot$
(Mateo et al. 1996; $L_{bol}\!\simeq\!2L_V$) and 
$\Delta{t}\!\simeq\!5\times10^5 \ yr$.
About 100 carbon stars are thus expected. The number is a lower limit,
because the full extent of the SDG has not been taken into account
and the bolometric luminosity has therefore been underestimated.
Moreover, the duration over which carbon stars can be observed is prolonged
through mass transfer from binary evolution. 
All together the expected number of SDG carbon stars might be about 10 times
larger than hitherto found.

\subsection{What about the RR Lyrae stars ?}
A significant amount of RR Lyrae stars have been found and 
have been used to constrain the distance towards the 
SDG, see Sect.~2.2. As mentioned in Sect.~1.3
the distance of the LPVs is consistent with the
RR Lyrae found in the same field (NS97).
The LPVs likely have an age comparable to the age of the carbon stars 
and a metallicity of \mbox{Z\muspc$\simeq$\muspc0.008}.
\hfill\break
The metallicity of the RR Lyrae stars are expected to be considerably 
smaller, say between \mbox{Z\muspc=\muspc0.0004\to0.001}.
The GB stars from the parent population appears to be absent in
Fig.~2a. This could be due to 
a bias in the selection of the GB stars,
confusion of an old, metal-poor population
with the GB from a 0.1\to1.0~Gyr population with 
\mbox{Z\muspc=\muspc0.008},
or the parent population of the RR Lyrae stars has a metallicity
close to \mbox{Z\muspc=\muspc0.008}.
In addition, it is not clear how the SDG RR Lyrae stars
are related to, or even might originated from
the four globular clusters located at the SDG distance
(Da Costa{\muspc\&\muspc}Armandroff 1995).
\hfill\break
The present situation is rather confusing. It is not clear
at all what the actual metallicity of the parent population of
the SDG RR Lyrae stars is.
A thorough analysis of deep CMDs, similar to those from 
Marconi et al. (1997), with a significant number
of stars near the main sequence turn-off is required
to determine the age and metallicity of the oldest
population in the SDG. 

\subsection{A twist of fate ?}

\subsubsection{Radial velocities}
The radial velocities of the `bulge' carbon stars would provide 
an independent verification of the photometric analysis presented here,
concerning membership of the SDG. Tyson\mbox{\muspc\&\muspc}Rich (1991)
determined the radial velocities for 33 stars from the ALRW91 
sample. Their radial velocities would support only for a small
number of stars the suggestion that the ALRW91 carbon stars are actually
located in the SDG, while the majority of the stars are moving towards
us and could be Bulge members.  
\par
The spectroscopic results related to the CN line strength and the NaD
equivalent width would support membership to the SDG combined
with a binary evolution scenario for these stars.
Membership of the Bulge does not make sense, because there is 
in that case no trace of a predecessor, i.e. a brighter population
of carbon stars (see Azzopardi 1994, Fig.~2). 
One could argue that an independent verification
of the radial velocities, with a zeropoint based on other stars
than the one template star ROA~153 used by Tyson\mbox{\muspc\&\muspc}Rich
(1991), might shed some light on the present contradictory results. 
\hfill\break
However, if both the photometric and spectroscopic analysis are correct
then the ALRW91 carbon stars are related to the SDG. 
Some stars move away from us in the same direction as the SDG,
while others move towards us, away from the SDG.
This does not necessarily imply that we observe the ongoing disruption of
the SDG. It is not clear if an encounter of the 
SDG with the galactic disc could result in the bifurcation of the
radial velocity distribution for stars formed or located in the tidal tail. 
The numerical simulations from Johnston et al. (1995,
their Fig.~10c and 10d) do show that the stars from 
a 1~Gyr old, tidal stream have a different radial velocity 
when observed at about 10\degr\ further along the orbit.
But the presence of a moving group is not obvious, since
its velocity distribution is not easily distinguished 
from halo or bulge stars. 

\subsubsection{Completeness limits}
An intriguing point is that in the LMC 
carbon stars have not been identified 
with \mbox{M$_K\!\la$\muspc\to6\mag5}, while 
they are found in some dwarf spheroidal galaxies.
The majority of the carbon stars were found from green or near-IR
grism surveys (see Azzopardi et al. 1985b 
and references cited therein). It is not unlikely that 
the limiting magnitude mentioned above mark 
the completeness boundary due to heavily crowding.
More carbon stars might be present, but they have not yet been identified.
If they are variable, then the databases from the 
various microlensing projects 
(see contributions in Ferlet et al. 1997) 
will prove to be an important and valuable asset. 
A massive spectroscopic study combined with a near-IR 
photometry of the LPVs might reveal 
the carbon stars below the completeness boundary of the grism survey.
Finding faint carbon stars in the LMC would
provide a significant contribution to understand the 
origin of the low luminosity SDG carbon stars.
\par
As an aside it is interesting that the colours of the ALRW91
\& the SDG carbon stars have comparable colours with those
found in other dwarf spheroidals. The question rises if this implies
that they have comparable age and metallicities. If this is the case:
a) where are the LPVs, and
b) how would that change the star formation and chemical evolution 
history\muspc?
As shown in Fig.~2b the blue giant branch sequence does not necessarily
imply only an old, metal-poor population which is indistinguishable
from a young, metal-richer giant branch. 

\section{Summary}
$\circ$\quad The ALRW91 carbon stars are likely not located in the Bulge, but 
are related to the SDG. 
\hfill\break
$\circ$\quad The ALRW91 carbon stars are not metal-rich.
\hfill\break
$\circ$\quad The metallicity is about \mbox{Z\muspc=\muspc0.008} 
with an age between 0.1\to1 Gyr.
\hfill\break
$\circ$\quad A significant number of the ALRW91 carbon stars 
are located below the tip of the red giant branch. 
\hfill\break
$\circ$\quad The origin of the carbon stars one magnitude below 
the red giant branch tip cannot be explained with the evolution
of a TP-AGB star. 
\hfill\break
$\bullet$\quad The faint carbon stars likely originated
through mass transfer from binary evolution. 

\acknowledgements{G.~Bertelli, A. Bressan 
and C.~Chiosi are acknowledged for the many
discussions and useful comments they made while the work was in progress.
M.A.T. Groenewegen is acknowledged for 
suggestions afterwards.
The referee M. Azzopardi is acknowledged for updating the author 
on the faint SMC and Fornax carbon stars.
P.~Marigo is 
thanked for patiently discussing with me the theoretical aspects 
of carbon star evolution.
The research of Y.K.~Ng is supported by TMR grant ERBFMRX-CT96-0086
from the European Community.}

\hyphenation{Slijk-huis}


\begin{thebibliography}{}
\bibitem [1996]{}
Alard C., 1996, ApJ 458, L17
\bibitem [1997]{}
Alcock C., Allsman R.A., Alves D.R., et al., 1997, ApJ 474, 217
\bibitem [1995] {}
Azzopardi M., 1994, in proceedings third ESO/CTIO workshop 
{\it `The Local Group: comparative and global properties'},
, A. Layden, R.C. Smith and J. Storm (eds.), 129
\bibitem [1985] {}
Azzopardi M., Lequeux J., Westerlund B.E., 1985a, A\&A 144, 388
\bibitem [1985] {}
Azzopardi M., Lequeux J., Rebeirot E., 1985b, A\&A 145, L4
\bibitem [1986] {}
Azzopardi M., Lequeux J., Westerlund B.E., 1986, A\&A 161, 232
\bibitem [1988] {}
Azzopardi M., Lequeux J., Robeirot E., 1988, A\&A 202, L27 (ALR88)
\bibitem [1991] {}
Azzopardi M., Lequeux J., Robeirot E., Westerlund B.E., 1991, A\&AS 88, 265
(ALRW91)
\bibitem [1997] {}
Azzopardi M., et al., 1997, private communication
%\bibitem [1991] {}
%Bersanelli M., Bouchet P., Falomo R., 1991, A\&A 252, 854

\bibitem [1994] {}
Bertelli G., Bressan A., Chiosi C., Fagotto F., Nasi E., 1994,
A\&AS 106, 275
\bibitem [1992]{Blommaert} Blommaert J.A.D.L., 1992,
Ph.D. thesis, Leiden University, the Netherlands 

\bibitem [1978] {}
Blanco V.M., Blanco B.M, McCarthy M.F., 1978, Nature 271, 638
\bibitem [1989] {}
Blanco V.M., Terndrup D.M., 1989, AJ 98, 843
\bibitem [1993] {}
Boothroyd A.I., Sackmann I.-J., Ahern S.C., 1993, ApJ 416, 762 
\bibitem [1991] {}
Bouchet P., Manfroid J., Schmider F.X., 1991,
A\&AS 91, 409
\bibitem []{}
Carter B.S., 1990, MNRAS 242, 1
\bibitem []{}
Chiosi C., Bertelli G., Bressan A., 1992, ARA\&A 30, 235
\bibitem []{}
Da Costa G., Armandroff T., 1995, AJ 109, 2533
\bibitem [1981] {}
Engels D., Sherwood W.A., Wamsteker W., Schultz G.V., 1981, A\&AS 45, 5
\bibitem [1996] {}
Fahlman G.G., Mandushev G., Richer H.B., et al., 1996, ApJ 459, L65
\bibitem []{}
Ferlet R., Maillard J.-P., B. Raban, 1997, 12th IAP Colloquium 
{\it `Variable stars and the astrophysical returns from microlensing surveys'},
Editions Fronti\'eres
\bibitem []{}
Frogel J.A., Persson S.E., Cohen J.G., 1980, ApJ 239, 495
\bibitem []{}
Geisler D. and Friel E.D., 1992, AJ 104, 128
\bibitem [1974]{Glass}
Glass I.S., 1974, MNASSA 33, 53, 71
\bibitem [1995]{Glass}
Glass I.S., Whitelock P.A., Catchpole R.M., Feast M.W., 1995, MNRAS 273, 383
\bibitem [1993]{}
Green P.J., 1997, proceedings IAU symposium 177, 
{\it`The carbon star phenomenon'}, R.F. Wing (ed.), {\it in press}
{\tt(=astro-ph/9608003)}
\bibitem [1993]{}
Groenewegen M.A.T., de Jong T., 1993, A\&A 267, 410
\bibitem [1995]{}
Groenewegen M.A.T., van den Hoek L.B., de Jong T., 1995, A\&A 293, 381
\bibitem [1996]{}
Groenewegen M.A.T., Whitelock P.A., 1996, MNRAS 281, 1347 (GW96)
\bibitem[1997]{}
Hron J., Kerschbaum F., Ng Y.K., Schultheis M., 1997,
A\&A {\it in preparation}
\bibitem[]{}
Ibata R., Gilmore G., Irwin M.J., 1994, Nature 370, 194 (IGI94)
\bibitem[]{}
Ibata R., Gilmore G., Irwin M.J., 1995, MNRAS 277, 781 (IGI95)
\bibitem[]{}
Ibata R., Wyse R.F.G., Gilmore G., Irwin M.J., Suntzeff N.B., 
1997, AJ 113, 635
\bibitem [1995]{}
Johnston K.V., Spergel D.N., Hernquist L., 1995, ApJ 451, 598
\bibitem [1983a] {}
Koornneef J., 1983a, A\&AS 51, 489
\bibitem [1983b] {}
Koornneef J., 1983b, A\&A 128, 84
\bibitem [1989]{}
Lattanzio J.C., 1989, ApJ 344, L25
\bibitem [1990]{}
Lequeux J., 1990, in proceedings {\it`From Miras to planetary nebulae:
which path for stellar evolution?'}, Montpellier (France), Sept.~4\to7,
1989, M.-O.~Mennessier and A. Omont (eds.), Editions Fronti\'eres, 273

\bibitem[]{}
Marconi G., Buonanno R., Castellani M., et al. 1997, A\&A {\it submitted}
({\tt =astro-ph/9703081})
\bibitem[]{}
Marigo P., Bressan A., Chiosi C., 1996a, A\&A 313, 545
\bibitem[]{}
Marigo P., Girardi L., Chiosi C., 1996b, A\&A 316, L1
\bibitem[]{}
Mateo M., Kubiak M., Szyma\'nski M., et al., 1995, AJ 110, 1141
\bibitem[]{}
Mateo M., Mirabel N., Udalski A., et al., 1996, ApJ 458, L13
\bibitem[]{}
McCarthy M.F., Blanco V.M., 1983, Mem.S.A.It. 54, 65
\bibitem[]{}
McClure R.D., 1984, ApJ 280, L31
\bibitem[]{}
McWilliam A., Rich R.M., 1994, ApJS 91, 749
\bibitem[]{}
Mendoza V. E.E., Johnson H.L., 1965, ApJ 141, 161
\bibitem[]{}
Ng Y.K., Bertelli, G., Bressan, A., Chiosi, C., Lub, J.,
1995, A\&A 295, 655 (erratum A\&A 301, 318) 
\bibitem[]{}
Ng Y.K., Bertelli G., Bressan A., Chiosi C., 1996, A\&A 310, 711
\bibitem [1997]{}
Ng Y.K., Schultheis M., 1997, A\&AS 123, 115
(NS97)%, {\tt =astro-ph/9609134})
\bibitem [1997]{}
Ng Y.K., Schultheis M., Hron J., 1997, A\&A {\it in preparation}

\bibitem [1994] {}
Paczy\'nski B., et al., 1994, AJ 107, 2060
%\bibitem [1993] {}
%Reid M.J., 1993, ARA\&A 31, 345
\bibitem [1975] {}
Reimers D., 1975, Mem. Soc. R. Li\`ege, Ser. 6(8), 369
\bibitem [1983] {}
Renzini A., Buzzoni A., 1983, Mem.S.A.It, 54, 335
\bibitem [1986] {}
Renzini A., Buzzoni A., 1986, in {\it`Spectral evolution of galaxies'},
C. Chiosi and A. Renzini (eds.), 135

\bibitem [1988] {}
Rich R.M., 1988, AJ 95, 828
\bibitem [1990] {}
Rich R.M., 1990, ApJ 362, 604
\bibitem [1985]{}
Rieke G.H., Lebofsky M.J., 1985, ApJ 288, 618
\bibitem [1996]{}
Sadler E.M., Rich R.M., Terndrup D.M., 1996, AJ 112, 171
\bibitem [1996]{}
Sarajedini A., Layden A.C., 1995, AJ 109, 1086 (SL95)
\bibitem [1996]{}
Schmecker-Hane T.A., Stetson P.B., Hesser J.E., Vandenberg D.A., 1996,
proceedings {\it `From stars to galaxies'}, ASP conference Series Vol.~98,
C.~Leitherer and U.~Fritze-von Alvensleben (eds.), 328
\bibitem [1997]{}
Schultheis M., Ng Y.K., Hron J., Kerschbaum F., 1997,
A\&A {\it submitted}
\bibitem []{}
Suntzeff N.B., Philips, M.M., Elias J.H., et al., 1993, PASP 105, 350
\bibitem []{}
Terndrup D.M., Frogel J.A., Whitford A.E., 1990, ApJ 357, 453
\bibitem []{}
Tyson N., Rich R.M., 1991, ApJ 367, 547
\bibitem [1995] {}
Vel\'azquez H., White S.D.M., 1995, MNRAS 275, L23 
\bibitem [1986] {}
Walker A.R., Mack P., 1986, MNRAS 220, 69
\bibitem [1981] {}
Wamsteker W., 1981, A\&A 97, 329
\hyphenation{Nijmegen}
\bibitem [1987] {}
Wesselink Th.J.H., 1987, Ph.D. thesis, Catholic University
Nijmegen, the Netherlands
\bibitem [1997] {}
Westerlund B.E., in {\it `The Magellanic Clouds'}, Cambridge 
Astrophysics Series 29, p7 \& p20 
\bibitem [1991] {}
Westerlund B.E., Lequeux J., Azzopardi M., Robeirot E., 1991, A\&A 244, 367
\bibitem [1995] {}
Westerlund B.E., Azzopardi M., Breysacher J., Robeirot E., 1993, A\&A 260, L4
\bibitem [1995] {}
Westerlund B.E., Azzopardi M., Breysacher J., Robeirot E., 1995, A\&A 303, 107
\bibitem [1996]{Whitelock}
Whitelock P.A., Irwin M., Catchpole R.M., 1996, New Astronomy 1, 57 (WIC96)
\bibitem [1983]{}
Whitford A.E., Rich R.M., 1983, ApJ 274, 723
\bibitem [1980]{}
Zinn R., 1980, ApJS 42, 19
\end{thebibliography}
\end{document}